\documentclass{svjour3}                     
\smartqed  
\usepackage{graphicx}
 \usepackage{mathptmx}      
\usepackage{amsmath}
\usepackage{latexsym}
\usepackage{amssymb}
\usepackage{dsfont}
\usepackage{color}
\usepackage{multirow}
\usepackage{enumitem}

\newcommand{\ud}{\mathrm{d}}

\newcommand{\uTr}{\mathrm{Tr}}

\newcommand{\uslash}{/\!\!\!\!}

\newcommand{\uS}{\vec{S}}

\newcommand{\uk}{\vec{k}}

\newcommand{\uzero}{\vec{0}}

\begin{document}

\title{Transverse-Momentum Distributions and Spherical Symmetry\thanks{Presented at the workshop "30 years of strong interactions", Spa, Belgium, 6-8 April 2011.}}


\author{C\'edric Lorc\'e         \and
        Barbara Pasquini 
}


\institute{C. Lorc\'e \at
              Institut fuer Kernphysik, Johannes Gutenberg-Universitaet,\\ 
              Mainz, D55099, Germany \\
              \email{lorce@kph.uni-mainz.de}           
           \and
              B. Pasquini \at
              Dipartimento di Fisica Nucleare e Teorica, Universit\`a degli Studi di Pavia, \\
              and INFN, Sezione di Pavia, I-27100 Pavia, Italy\\
              \email{pasquini@pv.infn.it}
}

\date{Received: date / Accepted: date}

\maketitle

\begin{abstract}
Transverse-momentum dependent parton distributions (TMDs) are studied in the framework of quark models. In particular, quark model relations among TMDs are reviewed and their physical origin is discussed in terms of rotational-symmetry properties of the nucleon state in its rest frame.
\keywords{Transverse-momentum distributions \and Relativistic quark models \and Spherical symmetry \and Light-cone helicity \and Canonical spin}
\PACS{12.39.-x \and 13.88.+e \and 13.60.-r}
\end{abstract}

\section{Introduction}
\label{intro}

Transverse-momentum dependent parton distributions (TMDs) have received a great attention in the last years as they represent key objects to map out the three-dimensional partonic structure of hadrons in momentum space. The dependence on the transverse momentum of the quark allows for non-trivial correlations between the orbital angular momentum and the spin of the quark inside nucleons with different polarization states. TMDs typically give rise to spin and azimuthal asymmetries in, for instance, semi-inclusive deep inelastic scattering and Drell-Yan processes, and significant efforts have already been devoted to measure these observables (see \emph{e.g.} Ref.~\cite{Barone:2010zz} for a recent review). However, the extraction of TMDs from experimental data is a quite difficult task and needs educated Ans\"atze for fits of TMD parametrizations. To this aim, model calculations of TMDs play a crucial role and are essential towards an understanding of the non-perturbative aspects of TMDs. 

Studies of the TMDs have been mainly focused on the quark contribution, and predictions have been obtained within a variety of models~\cite{Jakob:1997wg,Goldstein:2002vv,Pobylitsa:2002fr,Gamberg:2003ey,Lu:2004au,Cherednikov:2006zn,Lu:2006kt,Gamberg:2007wm,Pasquini:2009bv,Efremov:2009ze,She:2009jq,Zhu:2011zza,Lu:2010dt,Avakian:2009jt,Courtoy:2008dn,Courtoy:2009pc,Courtoy:2008vi,Pasquini:2010af,Brodsky:2010vs,Lorce:2007fa,Pasquini:2008ax,Boffi:2009sh,Pasquini:2010pa,Lorce:2011dv,Bacchetta:2008af,Avakian:2010br,Avakian:2008dz,Efremov:2010mt,Meissner:2007rx,Wakamatsu:2009fn,Schweitzer:2001sr,Bourrely:2010ng}. Despite the specific assumptions for modeling the quark dynamics, most of these models predicted relations among the leading-twist TMDs. Since in QCD these TMDs are all independent, it is clear that such relations should be traced back to some common simplifying assumptions in the models. First of all, it was noticed that they break down in models with gauge-field degrees of freedom. Furthermore, most quark models are valid at some very low scale and these relations are expected to break under QCD evolution to higher scales. Despite these limitations, such relations are intriguing because they can provide guidelines for building parametrizations of TMDs to be tested with experimental data and can also give useful insights for the understanding of the origin of the different spin-orbit correlations of quarks in the nucleon.

Here we discuss a straightforward derivation of the relations directly at the level of the amplitudes. This approach emphasizes the geometric origin of the relations since it does not rely on  specific assumptions for modeling the quark dynamics. In Sec.~\ref{sec:1} we quickly review the formalism for the definition of the leading-twist TMDs, and introduce a convenient representation of the quark-quark correlator in terms of the net-polarization states of the quark and the nucleon. The model relations among TMDs are then introduced in Sec.~\ref{sec:2}. In Sec.~\ref{sec:3} we identify the assumptions common to all models where the relations have been observed and derive the relations. 

An alternative derivation based on the language of wave functions can be found in Ref.~\cite{Lorce:2011zt}, where we also discuss an additional relation due to $SU(6)$ spin-flavor symmetry.

\section{Transverse-Momentum Dependent Parton Distributions}
\label{sec:1}

\subsection{Definitions}

In this section, we review the formalism for the definition of TMDs, following the conventions of Refs.~\cite{Mulders:1995dh,Boer:1997nt,Goeke:2005hb}. Introducing two lightlike four-vectors $n_\pm$ satisfying $n_+\cdot n_-=1$, we write the light-cone components of a generic four-vector $a$ as $\left[a^+,a^-,\vec a_\perp\right]$ with $a^\pm=a\cdot n_\mp$.
\newline
\noindent
The density of quarks can be defined from the following quark-quark correlator 
\begin{equation}
\Phi_{ab}(x,\uk_\perp,\uS)=\int\frac{\ud z^-\,\ud^2z_\perp}{(2\pi)^3}\,e^{i\left(k^+z^--\uk_\perp\cdot\vec z_\perp\right)}\langle P,\uS|\overline\psi_b(0)\mathcal U^{n_-}_{(0,+\infty)}\mathcal U^{n_-}_{(+\infty,z)}\psi_a(z)|P,\uS\rangle\big|_{z^+=0},
\label{correlator}
\end{equation}
where $k^+=xP^+$, $\psi$ is the quark field operator with $a,b$ indices in the Dirac space, and $\mathcal U$ is the Wilson line which ensures color gauge invariance \cite{Bomhof:2004aw}. The target state is characterized by its four-momentum $P$ and the direction of its polarization $\uS$. We choose a frame where the hadron momentum has no transverse components $P=\left[P^+,\tfrac{M^2}{2P^+},\uzero_\perp\right]$.

TMDs enter the general Lorentz-covariant decomposition of the correlator $\Phi_{ab}(x,\uk_\perp,\uS)$ which, at twist-two level and for a spin-$1/2$ target, reads
\begin{multline}
\Phi(x,\uk_\perp,\uS)=\frac{1}{2}\Big\{f_1\,\uslash n_+-\tfrac{\epsilon_T^{ij}\,k_\perp^iS_\perp^j}{M}\,f_{1T}^\perp\,\uslash n_++S_z\,g_{1L}\,\gamma_5\uslash n_++\tfrac{\uk_\perp\cdot\uS_\perp}{M}\,g_{1T}\,\gamma_5\uslash n_+\\
+h_{1T}\,\tfrac{[\uslash S_\perp,\uslash n_+]}{2}\,\gamma_5+S_z\,h_{1L}^\perp\,\tfrac{[\uslash k_\perp,\uslash n_+]}{2M}\,\gamma_5+\tfrac{\uk_\perp\cdot\uS_\perp}{M}\,h_{1T}^\perp\,\tfrac{[\uslash k_\perp,\uslash n_+]}{2M}\,\gamma_5+ih_1^\perp\,\tfrac{[\uslash k_\perp,\uslash n_+]}{2M}\Big\},
\end{multline}
where $\epsilon_T^{12}=-\epsilon_T^{21}=1$, and the transverse four-vectors are defined as $a_\perp=\left[0,0,\vec a_\perp\right]$. The nomenclature of the distribution functions follows closely that of Ref.~\cite{Mulders:1995dh}, sometimes referred to as ``Amsterdam notation''. Among these eight distributions, the so-called Boer-Mulders function $h_1^\perp$~\cite{Boer:1997nt} and Sivers function $f_{1T}^\perp$~\cite{Sivers:1989cc} are T-odd, \emph{i.e.} they change sign under ``naive time-reversal'', which is defined as usual time-reversal but without interchange of initial and final states. All the TMDs depend on $x$ and $\uk^2_\perp$. These functions can be individually isolated by performing traces of the correlator with suitable Dirac matrices. Using the abbreviation $\Phi^{[\Gamma]}\equiv\uTr[\Phi\Gamma]/2$, we have
\begin{subequations}
\begin{align}
\Phi^{[\gamma^+]}(x,\uk_\perp,\uS)&=f_1-\tfrac{\epsilon_T^{ij}\,k_\perp^iS_\perp^j}{M}\,f_{1T}^\perp,
\label{vector}\\
\Phi^{[\gamma^+\gamma_5]}(x,\uk_\perp,\uS)&=S_z\,g_{1L}+\tfrac{\uk_\perp\cdot\uS_\perp}{M}\,g_{1T},\\
\Phi^{[i\sigma^{j+}\gamma_5]}(x,\uk_\perp,\uS)&=S_\perp^j\,h_1+S_z\,\tfrac{k_\perp^j}{M}\,h^\perp_{1L}+S^i_\perp\,\tfrac{2k^i_\perp k^j_\perp-\uk^2_\perp\delta^{ij}}{2M^2}\,h^\perp_{1T}+\tfrac{\epsilon_T^{ji}\,k_\perp^i}{M}\,h_1^\perp,
\label{tensor}
\end{align}
\end{subequations}
where $j=1,2$ is a transverse index, and $h_1=h_{1T}+\tfrac{\uk^2_\perp}{2M^2}\,h^\perp_{1T}$.

The correlation function $\Phi^{[\gamma^+]}(x,\uk_\perp,\uS)$ is just the unpolarized quark distribution, which integrated over $\uk_\perp$ gives the familiar light-cone momentum distribution $f_1(x)$. All the other TMDs characterize the strength of different spin-spin and spin-orbit correlations. The precise form of this correlation is given by the prefactors of the TMDs in Eqs.~\eqref{vector}-\eqref{tensor}. In particular, the TMDs $g_{1L}$ and $h_1$ describe the strength of a correlation between a longitudinal/transverse target polarization and a longitudinal/transverse parton polarization. After integration over $\uk_\perp$, they reduce to the helicity and transversity distributions, respectively. By definition, the spin-orbit correlations described by $f_{1T}^\perp$, $g_{1T}$, $h_1^\perp$, $h_{1L}^\perp$ and $h_{1T}^\perp$ involve the transverse parton momentum and the polarization of both the parton and the target, and vanish upon integration over $\uk_\perp$.

In the following we will focus the discussion on the quark contribution to TMDs, ignoring the contribution from gauge fields and therefore reducing the gauge links in Eq.~(\ref{correlator}) to the identity.

\subsection{Net-Polarization Basis}
\label{sec:12}

The physical meaning of the correlations encoded in TMDs becomes especially transparent when expressed in the basis of net polarization for the quark and the nucleon, see Refs.~\cite{Lorce:2011dv,Lorce:2011zt}. In this basis, the correlator \eqref{correlator} for a given quark flavor $q$ can be written in a matrix form 
\begin{equation}
\Phi^{\mu\nu}_q=\begin{pmatrix}
f^q_1&\tfrac{k_\perp}{M}\,h_1^{\perp q}&0&0\\
\tfrac{k_\perp}{M}\,f_{1T}^{\perp q}&h_{1T}^{-q}&0&0\\
0&0&h_{1T}^{+q}&\tfrac{k_\perp}{M}\,g_{1T}^q\\
0&0&\tfrac{k_\perp}{M}\,h_{1L}^{\perp q}&g_{1L}^q
\end{pmatrix},\label{tensor2}
\end{equation}
where we introduced the notations $h_{1T}^{\pm q}=h^q_1\pm\tfrac{\uk_\perp^2}{2M^2}\,h_{1T}^{\perp q}$ and chose for convenience the axes in the transverse plane such that $\vec k_\perp=k_\perp\,\vec e_y$. The four-component index\footnote{Note this is \emph{not} a Lorentz index but Einstein's summation convention still applies.} $\mu$ refers to the net polarization of the nucleon: $\mu=0$ stands for an unpolarized $\tfrac{1}{2}(\uparrow+\downarrow)$ nucleon, while $\mu=i=1,2,3$ stands for a nucleon with net polarization $\tfrac{1}{2}(\uparrow-\downarrow)$ in the $i$th direction. Likewise, $\nu=0$ stands for an unpolarized quark $(\Gamma=\gamma^+)$, $\nu=j=1,2$ stands for a quark with net polarization in the $j$th direction $(\Gamma=i\sigma^{j+}\gamma_5)$, and $\nu=3$ stands for a quark with net polarization in the $z$-direction $(\Gamma=\gamma^+\gamma_5)$.

\section{Model Relations}
\label{sec:2}

In QCD, the eight TMDs are all independent. It appeared however in a large panel of low-energy quark models that relations among some TMDs exist. At twist-two level, there are three flavor-independent relations\footnote{Other expressions can be found in the literature, but are just combinations of the relations \eqref{rel1}-\eqref{rel3}.}, two are linear and one is quadratic in the TMDs
\begin{gather}
g^q_{1L}-\left[h^q_1+\tfrac{\uk_\perp^2}{2M^2}\,h_{1T}^{\perp q}\right]=0,\label{rel1}\\
g^q_{1T}+h_{1L}^{\perp q}=0,\label{rel2}\\
\left(g^q_{1T}\right)^2+2h^q_1\,h_{1T}^{\perp q}=0.\label{rel3}
\end{gather}
A further flavor-dependent relation involves both polarized and unpolarized TMDs
\begin{equation}\label{rel4}
\mathcal D^qf_1^q+g_{1L}^q=2h_1^q,
\end{equation}
where, for a proton target, the flavor factors with $q=u,d$ are given by $\mathcal D^u=\tfrac{2}{3}$ and $\mathcal D^d=-\tfrac{1}{3}$. As discussed in Ref.~\cite{Avakian:2010br}, at variance with the relations \eqref{rel1}-\eqref{rel3}, the flavor dependence in the relation \eqref{rel4} requires specific assumptions for the spin-isospin structure of the nucleon state, like $SU(6)$ spin-flavor symmetry. In this proceeding we focus only on the flavor-independent relations \eqref{rel1}-\eqref{rel3}. A discussion of the flavor-dependent one can be found in \cite{Lorce:2011zt}.

The interest in these relations is purely phenomenological. In order to interpret the experimental data sensitive to TMDs, one needs inputs from educated models and parametrizations for the extraction of these distributions. It is therefore particularly interesting to see to what extent the relations \eqref{rel1}-\eqref{rel4} can be useful as \emph{approximate} relations, which provide simplified and intuitive notions for the interpretation of the data. Note that some preliminary calculations in lattice QCD give indications that the relation \eqref{rel2} may indeed be approximately satisfied \cite{Hagler:2009mb,Musch:2010ka}. A discussion on how general these relations are can be found in Ref.~\cite{Avakian:2010br}. Let us just mention that they were observed in the bag model \cite{Avakian:2010br,Avakian:2008dz}, light-cone constituent quark models \cite{Pasquini:2008ax}, some quark-diquark models \cite{Jakob:1997wg,She:2009jq,Zhu:2011zza}, the covariant parton model \cite{Efremov:2009ze} and more recently in the light-cone version of the chiral quark-soliton model \cite{Lorce:2011dv}. Note however that there also exist models where the relations are not satisfied, like in some versions of the spectator model~\cite{Bacchetta:2008af} and the quark-target model \cite{Meissner:2007rx}. 

\section{Derivation of the Flavor-Independent Relations}
\label{sec:3}

We show in this section that the flavor-independent relations \eqref{rel1}-\eqref{rel3} can easily be derived, once the following assumptions are made:
\begin{enumerate}
\item the probed quark behaves as if it does not interact directly with the other partons ({\em i.e.} one works within the standard impulse approximation) and there are no explicit gluons;
\item the quark light-cone and canonical polarizations are related by a rotation with axis orthogonal to both $\uk_\perp$ and the light-cone direction;
\item the target has spherical symmetry in the canonical-spin basis.
\end{enumerate}
From these assumptions, one realizes that the flavor-independent relations have essentially a geometrical origin, as was already guessed in the context of the bag model almost a decade ago \cite{Efremov:2002qh}. We note however that the spherical symmetry is a sufficient but not necessary condition for the validity of the individual flavor-independent relations. The assumptions 1.-3. are satisfied by all models where the relations have been observed and can be considered as necessary conditions \cite{Lorce:2011zt}. Consistently, the few known models where the relations are absent \cite{Bacchetta:2008af,Meissner:2007rx} fail with at least one of the above three conditions.

As we have seen, the TMDs can be expressed in simple terms using light-cone polarization. On the other hand, rotational symmetry is easier to handle in terms of canonical polarization, which is the natural one in the instant form. We therefore write the TMDs in the canonical-spin basis, and then impose spherical symmetry. But before that, we need to know how to connect light-cone helicity to canonical spin.

\subsection{Connection between Light-Cone Helicity and Canonical Spin}

Relating in general light-cone helicity with canonical spin is usually quite complicated, as the dynamics is involved. Fortunately, the common approach in quark models is to assume that the target can be described by quarks without mutual interactions. In this case the connection simply reduces to a rotation in polarization space with axis orthogonal to both $\uk_\perp$ and $\vec e_z$. The quark creation operator with canonical spin $\sigma$ can then be written in terms of quark creation operators with light-cone helicity $\lambda$ as follows
\begin{equation}\label{genMelosh}
q^\dag_\sigma=\sum_\lambda D^{(1/2)*}_{\sigma\lambda}\,q^\dag_\lambda\qquad\text{with}\qquad D^{(1/2)*}_{\sigma\lambda}=\begin{pmatrix}\cos\tfrac{\theta}{2}&-\hat k_R\,\sin\tfrac{\theta}{2}\\\hat k_L\,\sin\tfrac{\theta}{2}&\cos\tfrac{\theta}{2}\end{pmatrix},
\end{equation}
where $\hat k_{R,L}=(k_x\pm ik_y)/k_\perp$. Note that the rotation does not depend on the quark flavor. The angle $\theta$ between light-cone and canonical polarizations is usually a complicated function of the quark momentum $k$ and is specific to each model. It contains part of the model dynamics. The only general property is that $\theta\to 0$ as $k_\perp\to 0$. Due to our choice of reference frame where the target has no transverse momentum, the light-cone helicity and canonical spin of the target can be identified, at variance with the quark polarizations.

\subsection{TMDs in Canonical-Spin Basis}

The four-component notation introduced in Sec.~\ref{sec:12} is very convenient for discussing the rotation between canonical spin and light-cone helicity at the amplitude level. One can easily see that the canonical tensor correlator $\Phi^{\mu\nu}_{Cq}$ is related to the light-cone one in Eq.~\eqref{tensor2} as follows
\begin{equation}\label{expectation}
\Phi^{\mu\nu}_{Cq}=\Phi_q^{\mu\rho}\,O_\rho^{\phantom{\rho}\nu},
\end{equation}
with the orthogonal matrix $O$, representing the rotation at the amplitude level, given by (remember that we chose $\vec k_\perp=k_\perp\,\vec e_y$)
\begin{equation}
O_\rho^{\phantom{\rho}\nu}=\begin{pmatrix}
\phantom{0}1\phantom{0}&0&0&0\\
0&\phantom{0}1\phantom{0}&0&0\\
0&0&\cos\theta&-\sin\theta\\
0&0&\sin\theta&\cos\theta
\end{pmatrix}.
\end{equation}
The canonical tensor correlator then takes the form
\begin{equation}\label{Ctensor}
\Phi^{\mu\nu}_{Cq}=\begin{pmatrix}
f^q_1&\frac{k_\perp}{M}\,h^{\perp q}_1&0&0\\
\frac{k_\perp}{M}\,f^{\perp q}_{1T}&h_{1T}^{-q}&0&0\\
0&0&\mathfrak h_{1T}^{+q}&\frac{k_\perp}{M}\,\mathfrak g_{1T}^q\\
0&0&\frac{k_\perp}{M}\,\mathfrak h^{\perp q}_{1L}&\mathfrak g^q_{1L}
\end{pmatrix},
\end{equation}
where we introduced the notations
\begin{align}
\begin{pmatrix}
\mathfrak g^q_{1L}\\ \tfrac{k_\perp}{M}\,\mathfrak h_{1L}^{\perp q}
\end{pmatrix}
&=\begin{pmatrix}\cos\theta&-\sin\theta\\\sin\theta&\cos\theta\end{pmatrix}\begin{pmatrix}
g^q_{1L}\\ \tfrac{k_\perp}{M}\,h_{1L}^{\perp q}
\end{pmatrix},\label{rot1}\\
\begin{pmatrix}
\tfrac{k_\perp}{M}\,\mathfrak g^q_{1T}\\ \mathfrak h_{1T}^{+q}
\end{pmatrix}
&=\begin{pmatrix}\cos\theta&-\sin\theta\\\sin\theta&\cos\theta\end{pmatrix}\begin{pmatrix}
\tfrac{k_\perp}{M}\,g^q_{1T}\\ h_{1T}^{+q}
\end{pmatrix}.\label{rot2}
\end{align}
Comparing Eq.~\eqref{Ctensor} with Eq.~\eqref{tensor2}, we observe that the multipole structure is conserved under the rotation~\eqref{expectation}. The rotation from light-cone to canonical polarizations affects only some of the multipole magnitudes, see Eqs.~\eqref{rot1} and \eqref{rot2}.

\subsection{Spherical Symmetry}

\begin{figure}[t!]
\begin{center}
\includegraphics[width=12cm]{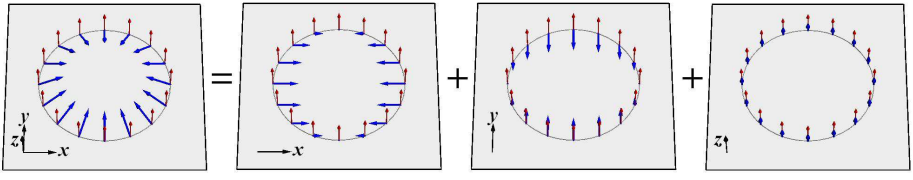}
\end{center}
\caption{\footnotesize{Net light-cone polarization (thick blue arrows) associated to a quark with net longitudinal canonical polarization (thin red arrows), and its vector decomposition along the three axes, for fixed $x$ and $k_\perp$ but arbitrary direction $\hat k_\perp$. The $x$- and $y$-components are pure dipoles, while the $z$-component is a pure monopole.}}
\label{fig1}
\end{figure}

\begin{figure}[t!]
\begin{center}
\includegraphics[width=12cm]{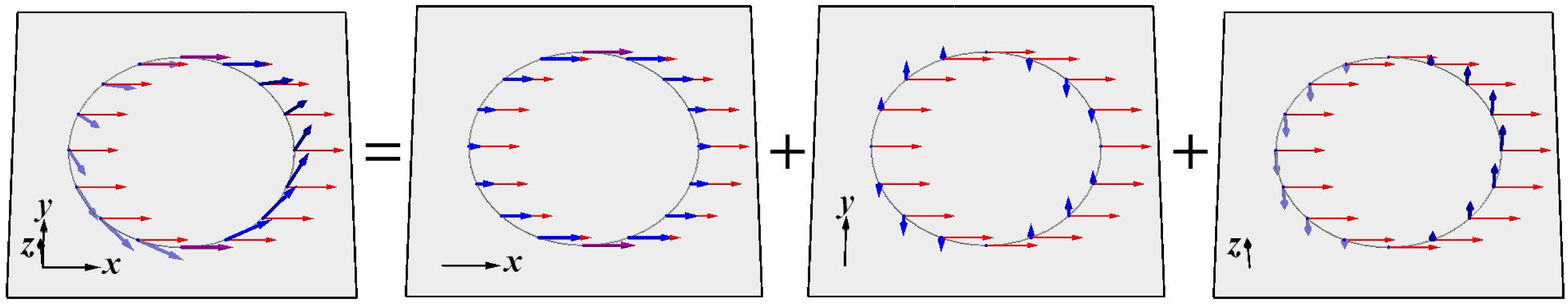}\vspace{.75cm}
\includegraphics[width=9cm]{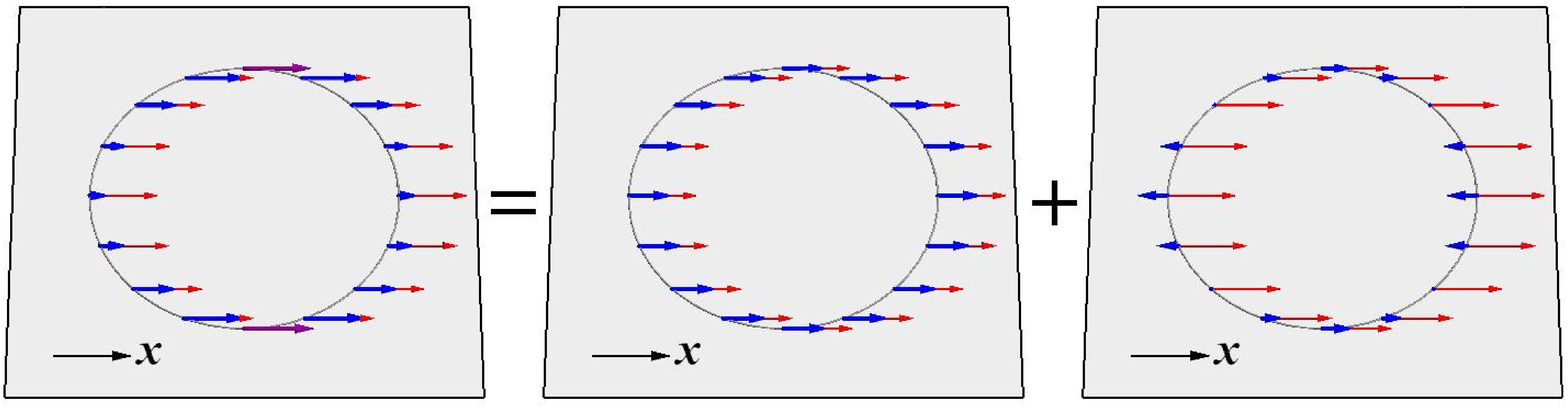}
\end{center}
\caption{\footnotesize{In the first line is shown the net light-cone polarization (thick blue arrows) associated to a quark with net canonical polarization in the $x$-direction (thin red arrows), and its vector decomposition along the three axes, for fixed $x$ and $k_\perp$ but arbitrary direction $\hat k_\perp$. The $y$-component is a pure quadrupole, and the $z$-component is a pure dipole.
The $x$-component is the sum of a monopole and a quadrupole, as 
illustrated in the second line.}}
\label{fig2}
\end{figure}

We are now ready to discuss the implications of spherical symmetry in the canonical-spin basis. Spherical symmetry means that the canonical tensor correlator has to be invariant  $O_R^T\Phi_{Cq}O_R=\Phi_{Cq}$ under any spatial rotation $O_R=\left(\begin{smallmatrix}1&0\\0&R\end{smallmatrix}\right)$ with $R$ the ordinary $3\times 3$ rotation matrix. It is equivalent to the statement that the tensor correlator has to commute with all the elements of the rotation group $\Phi_{Cq}O_R=O_R\Phi_{Cq}$. As a result of Schur's lemma, the canonical tensor correlator must have the following structure
\begin{equation}
\Phi^{\mu\nu}_{Cq}=\begin{pmatrix}A^q&0&0&0\\0&B^q&0&0\\0&0&B^q&0\\0&0&0&B^q\end{pmatrix}.
\end{equation}
Comparing this with Eq.~\eqref{Ctensor}, we conclude that spherical symmetry implies
\begin{subequations}
\begin{gather}
f^q_1=A^q,\label{constraint0}\\
\mathfrak g_{1L}^q=\mathfrak h_{1T}^{+q}=h_{1T}^{-q}=B^q,\label{constraint1}\\
\mathfrak g_{1T}^q=\mathfrak h_{1L}^{\perp q}=f_{1T}^{\perp q}=h_1^{\perp q}=0.\label{constraint2}
\end{gather}
\end{subequations}
Spherical symmetry in the canonical-spin basis implies that only the monopole structures have non-vanishing amplitude in the canonical tensor correlator $\Phi_{Cq}$. Note however that because of the rotation connecting canonical and light-cone polarizations, see Eq.~\eqref{expectation}, higher multipole amplitudes are non-vanishing in the tensor correlator $\Phi_q$, as illustrated in Figs.~\ref{fig1} and \ref{fig2}. It follows that spherical symmetry imposes some relations among the multipole structures in the light-cone helicity basis, and therefore among the TMDs. Inserting the constraints \eqref{constraint1} and \eqref{constraint2} into Eqs.~\eqref{rot1} and \eqref{rot2}, we automatically obtain the flavor-independent relations \eqref{rel1}-\eqref{rel3}.

\section{Conclusions}

In this work we presented a study of the transverse-momentum dependent parton distributions in the framework of quark models. We focused the discussion on model relations which appear in a large panel of quark models, elucidating their physical origin and implications. We have shown that these model relations have essentially a geometrical origin, and can be traced back to properties of rotational invariance of the system. In particular, we identified the conditions which are sufficient for the existence of the flavor-independent relations.

We presented a derivation of the relations based on the representation of the quark correlator entering the definition of TMDs in terms of the polarization amplitudes of the quarks and nucleon. Such amplitudes are usually expressed in the basis of light-cone helicity. However, in order to discuss in a simple way the rotational properties of the system, we introduced the representation in the basis of canonical spin and showed how both basis are related.

Finally, we remark that the model relations are not expected to hold identically in QCD where TMDs are all independent. However, they provide simplified and intuitive notions for the interpretation of the spin and orbital angular momentum structure of the nucleon. As such, they can be useful for phenomenological studies to build up simplified parametrizations of TMDs to be fitted to data. Furthermore, the comparison with the experimental data will tell us the degree of accuracy of such relations, giving insights for further studies towards more refined quark models.

\begin{acknowledgements}
This work was supported in part by the Research Infrastructure Integrating Activity ``Study of Strongly Interacting Matter'' (acronym HadronPhysics2, Grant Agreement n. 227431) under the Seventh Framework Programme of the European Community, by the Italian MIUR through the PRIN 2008EKLACK ``Structure of the nucleon: transverse momentum, transverse spin and orbital angular momentum''.
\end{acknowledgements}


\begin{thebibliography}{}
%
%

\bibitem{Barone:2010zz}
  V.~Barone, F.~Bradamante and A.~Martin,
  Prog.\ Part.\ Nucl.\ Phys.\  {\bf 65}, 267 (2010).

\bibitem{Jakob:1997wg}
  R.~Jakob, P.~J.~Mulders and J.~Rodrigues,
  Nucl.\ Phys.\  A {\bf 626}, 937 (1997).

\bibitem{Goldstein:2002vv}
  G.~R.~Goldstein and L.~Gamberg,
  arXiv:hep-ph/0209085.

\bibitem{Pobylitsa:2002fr}
  P.~V.~Pobylitsa,
  arXiv:hep-ph/0212027.

\bibitem{Gamberg:2003ey}
  L.~P.~Gamberg, G.~R.~Goldstein and K.~A.~Oganessyan,
  Phys.\ Rev.\  D {\bf 67}, 071504 (2003).

\bibitem{Lu:2004au}
  Z.~Lu and B.~Q.~Ma,
  Nucl.\ Phys.\  A {\bf 741}, 200 (2004).

\bibitem{Cherednikov:2006zn}
  I.~O.~Cherednikov, U.~D'Alesio, N.~I.~Kochelev and F.~Murgia,
  Phys.\ Lett.\  B {\bf 642}, 39 (2006).

\bibitem{Lu:2006kt}
  Z.~Lu and I.~Schmidt,
  Phys.\ Rev.\  D {\bf 75}, 073008 (2007).

\bibitem{Gamberg:2007wm}
  L.~P.~Gamberg, G.~R.~Goldstein and M.~Schlegel,
  Phys.\ Rev.\  D {\bf 77}, 094016 (2008).

\bibitem{Pasquini:2009bv}
  B.~Pasquini, S.~Boffi and P.~Schweitzer,
  Mod.\ Phys.\ Lett.\  A {\bf 24}, 2903 (2009).

\bibitem{Efremov:2009ze}
  A.~V.~Efremov, P.~Schweitzer, O.~V.~Teryaev and P.~Zavada,
  Phys.\ Rev.\  D {\bf 80}, 014021 (2009).

\bibitem{She:2009jq}
  J.~She, J.~Zhu and B.~Q.~Ma,
  Phys.\ Rev.\  D {\bf 79}, 054008 (2009).

\bibitem{Zhu:2011zza}
  J.~Zhu and B.~Q.~Ma,
  Phys.\ Lett.\  B {\bf 696}, 246 (2011).

\bibitem{Lu:2010dt}
  Z.~Lu and I.~Schmidt,
  Phys.\ Rev.\  D {\bf 82}, 094005 (2010).

\bibitem{Avakian:2009jt}
  H.~Avakian, A.~V.~Efremov, P.~Schweitzer, O.~V.~Teryaev, F.~Yuan and P.~Zavada,
  Mod.\ Phys.\ Lett.\  A {\bf 24}, 2995 (2009).

\bibitem{Courtoy:2008dn}
  A.~Courtoy, S.~Scopetta and V.~Vento,
  Phys.\ Rev.\  D {\bf 79}, 074001 (2009).

\bibitem{Courtoy:2009pc}
  A.~Courtoy, S.~Scopetta and V.~Vento,
  Phys.\ Rev.\  D {\bf 80}, 074032 (2009).

\bibitem{Courtoy:2008vi}
  A.~Courtoy, F.~Fratini, S.~Scopetta and V.~Vento,
  Phys.\ Rev.\  D {\bf 78}, 034002 (2008).

\bibitem{Pasquini:2010af}
  B.~Pasquini and F.~Yuan,
  Phys.\ Rev.\  D {\bf 81}, 114013 (2010).

\bibitem{Brodsky:2010vs}
  S.~J.~Brodsky, B.~Pasquini, B.~W.~Xiao and F.~Yuan,
  Phys.\ Lett.\  B {\bf 687}, 327 (2010).

\bibitem{Lorce:2007fa}
  C.~Lorc\'e,
  Phys.\ Rev.\  D {\bf 79}, 074027 (2009). 

\bibitem{Pasquini:2008ax}
  B.~Pasquini, S.~Cazzaniga and S.~Boffi,
  Phys.\ Rev.\  D {\bf 78}, 034025 (2008).

\bibitem{Boffi:2009sh}
  S.~Boffi, A.~V.~Efremov, B.~Pasquini and P.~Schweitzer,
  Phys.\ Rev.\  D {\bf 79}, 094012 (2009).

\bibitem{Pasquini:2010pa}
  B.~Pasquini and C.~Lorc\'e,
  arXiv:1008.0945 [hep-ph].

\bibitem{Lorce:2011dv}
  C.~Lorc\'e, B.~Pasquini and M.~Vanderhaeghen,
  JHEP {\bf 1105}, 041 (2011).

\bibitem{Bacchetta:2008af}
  A.~Bacchetta, F.~Conti and M.~Radici,
  Phys.\ Rev.\  D {\bf 78}, 074010 (2008).

\bibitem{Avakian:2010br}
  H.~Avakian, A.~V.~Efremov, P.~Schweitzer and F.~Yuan,
  Phys.\ Rev.\  D {\bf 81},  074035 (2010).

\bibitem{Avakian:2008dz}
  H.~Avakian, A.~V.~Efremov, P.~Schweitzer and F.~Yuan,
  Phys.\ Rev.\  D {\bf 78}, 114024 (2008).

\bibitem{Efremov:2010mt}
  A.~V.~Efremov, P.~Schweitzer, O.~V.~Teryaev and P.~Zavada,
  arXiv:1012.5296 [hep-ph].

\bibitem{Meissner:2007rx}
  S.~Meissner, A.~Metz and K.~Goeke,
  Phys.\ Rev.\  D {\bf 76}, 034002 (2007).
  
\bibitem{Wakamatsu:2009fn}
  M.~Wakamatsu,
  Phys.\ Rev.\  D {\bf 79}, 094028 (2009).

\bibitem{Schweitzer:2001sr}
  P.~Schweitzer, D.~Urbano, M.~V.~Polyakov, C.~Weiss, P.~V.~Pobylitsa
  and K.~Goeke,
  Phys.\ Rev. D {\bf 64}, 034013 (2001).
  
\bibitem{Bourrely:2010ng}
  C.~Bourrely, F.~Buccella and J.~Soffer,
  Phys.\ Rev.\  D {\bf 83}, 074008 (2011).

\bibitem{Lorce:2011zt}
  C.~Lorc\'e and B.~Pasquini,
  arXiv:1104.5651 [hep-ph], to appear in Phys. Rev. D.

\bibitem{Mulders:1995dh}
  P.~J.~Mulders and R.~D.~Tangerman,
  Nucl.\ Phys.\  B {\bf 461}, 197 (1996) 
  [Erratum-ibid.\  B {\bf 484}, 538 (1997)].
  
\bibitem{Boer:1997nt}
  D.~Boer and P.~J.~Mulders,
  Phys.\ Rev.\  D {\bf 57}, 5780 (1998).
  
\bibitem{Goeke:2005hb}
  K.~Goeke, A.~Metz and M.~Schlegel,
  Phys.\ Lett.\  B {\bf 618}, 90 (2005).

\bibitem{Bomhof:2004aw}
  C.~J.~Bomhof, P.~J.~Mulders and F.~Pijlman,
  Phys.\ Lett.\  B {\bf 596}, 277 (2004).

\bibitem{Sivers:1989cc}
  D.~W.~Sivers,
  Phys.\ Rev.\  D {\bf 41}, 83 (1990).

\bibitem{Hagler:2009mb}
  Ph.~H\"agler, B.~U.~Musch, J.~W.~Negele and A.~Sch\"afer,
  Europhys.\ Lett.\  {\bf 88}, 61001 (2009).

\bibitem{Musch:2010ka}
  B.~U.~Musch, Ph.~H\"agler, J.~W.~Negele and A.~Sch\"afer,
  Phys.\ Rev.\  D {\bf 83}, 094507 (2011).
  
\bibitem{Efremov:2002qh}
  A.~V.~Efremov and P.~Schweitzer,
  JHEP {\bf 0308}, 006 (2003).



\end{thebibliography}


\end{document}